\documentclass[RNAAS]{aastex631}

\shorttitle{Gaia22eor: Eclipse of V2756~Sgr}
\shortauthors{Merc et al.}
\graphicspath{{./}{figures/}}

\begin{document}

\title{Gaia22eor: Symbiotic star V2756~Sgr detected in a deep eclipse}

\correspondingauthor{Jaroslav Merc}
\email{jaroslav.merc@mff.cuni.cz}

\author[0000-0001-6355-2468]{Jaroslav Merc}
\affiliation{Astronomical Institute, Faculty of Mathematics and Physics, Charles University\\V Hole\v{s}ovi\v{c}k{\'a}ch 2, 180 00 Prague, Czech Republic}






\author{Hamish Barker}
\affiliation{Rutherford Street Observatory, Nelson, New Zealand}
\affiliation{Variable Stars South Group}

\author[0000-0003-4299-6419]{Rudolf Gális}
\affiliation{Institute of Physics, Faculty of Science, P. J. \v{S}af{\'a}rik University\\Park Angelinum 9, 040 01 Ko\v{s}ice, Slovak Republic}


\begin{abstract}
V2756~Sgr is a long-known S-type symbiotic binary that was recently detected in $\sim$ 1 mag eclipse-like fading by the \textit{Gaia} satellite. This behavior was reported as a Gaia Science Alert under the designation Gaia22eor. V2756~Sgr has not been reported as an eclipsing symbiotic system. In this contribution, we have investigated the recent light curves of this target obtained by \textit{Gaia} and ASAS-SN survey and supplemented these data by the photometry from the ASAS survey and DASCH archive. In addition, low-resolution BP/RP spectra from \textit{Gaia} were examined. Based on the presented analysis, we conclude that the observed fading is indeed an eclipse of the hot component of V2756~Sgr by the cool giant. The data also allowed us to refine the orbital period to 725 $\pm$ 3 days. 
\end{abstract}



\section{Introduction} \label{sec:intro}

V2756~Sgr (=AS 293; $\alpha_{2000}$ = 18:14:34.52, \mbox{$\delta_{2000}$ = -29:49:23.99}) was known as an emission-line star since 1950s \citep{1950ApJ...112...72M}. Later, \citet{1969PNAS...63.1045H} classified it as a possible symbiotic system. He noted a strong continuum spectrum, prominent Balmer lines, and \ion{He}{2} 4686 \AA\,\,in emission. The spectrum presented by \citet{1984PASA....5..369A} showed an M-type continuum and emission lines with ionization potential up to 114 eV (Raman-scattered \ion{O}{6} lines), undoubtedly confirming the symbiotic nature of V2756~Sgr. 


Our attention was drawn to this object when it was detected in $\sim$ 1 mag fading (in $G$ filter) by the \textit{Gaia} satellite. This behavior was reported in the form of the Gaia Science Alert \citep[GSA;][]{2021A&A...652A..76H} on November 14, 2022 (under the designation Gaia22eor). In this contribution, we present the investigation of the light curves of V2756~Sgr from the ASAS survey \citep{1997AcA....47..467P}, the ASAS-SN survey \citep{2014ApJ...788...48S, 2017PASP..129j4502K}, the DASCH 
archive of digitized glass photographic plates of the Harvard College Observatory \citep{2010AJ....140.1062L}, and of the data obtained by the \textit{Gaia} satellite 
 that are available at the GSA webpage\footnote{http://gsaweb.ast.cam.ac.uk/alerts/alert/Gaia22eor/}. 

\begin{figure}[t]
\begin{center}
\includegraphics[scale=0.75,angle=0]{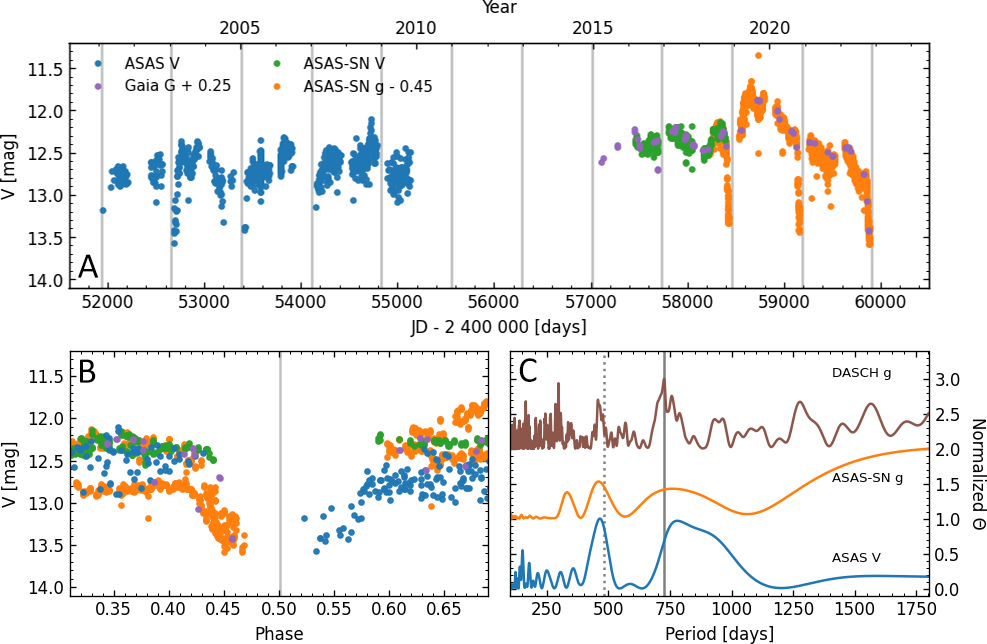}
\caption{Photometry of V2756~Sgr. \textbf{Panel A:} Long-term light curve of V2756~Sgr covering the period of the years 2001 - 2022, constructed on the basis of data from ASAS
, ASAS-SN
, and \textit{Gaia} 
. 
Vertical gray lines show the eclipse times. \textbf{Panel B: }Part of the light curve of V2756~Sgr phased with $P$ = 725\,days showing the observed eclipses. 
\textbf{Panel C:} Periodograms of the data from selected surveys.
The vertical solid and dotted lines denote the periods of 725 days (orbital) and 480 days \citep[given by][]{2013AcA....63..405G}, respectively. \label{fig:figure}}
\end{center}
\end{figure}

\section{Photometric behavior} \label{sec:photo}
Photometric observations of V2756~Sgr were first examined by \citet{1970IBVS..469....1H}. She analyzed $\sim$ 350 plates obtained at the Maria Mitchell Observatory and claimed that the star varies with a period of 243 days. More recently, \citet{2013AcA....63..405G} analyzed the light curve of V2756~Sgr obtained by the ASAS survey and suggested the orbital period of this symbiotic binary to be 480 $\pm$ 19 days.

\textit{Gaia} observed V2756~Sgr since March 2015. Some degree of variability has been well visible in the light curve since the beginning of observations (Fig.~\ref{fig:figure}A). However, the most prominent change is the recent (2022) fading of $>$ 1 mag in the $G$ filter (reported as a GSA). No previous deep fadings are seen in this dataset. 

Given very long gaps in the \textit{Gaia} observations, we supplemented these data with the observations obtained by the ASAS-SN survey. The light curve in the $g$ filter (Fig.~\ref{fig:figure}A) revealed two other similar events (observed in 2018 and 2020) at a time when no \textit{Gaia} observations were available. We attributed these brightness drops to the eclipses between the components of the symbiotic system. Using the eclipses, we obtained the orbital period of V2756~Sgr, which is 725 $\pm$ 3 days. The timescale is similar to the orbital variability observed in other S-type symbiotic binaries \citep[e.g.,][]{2012BaltA..21....5M, 2013AcA....63..405G}. The estimated duration of the eclipse is $\sim$\,115 days.

No fadings are seen in the ASAS-SN $V$ observations. It turned out that this was just due to an unfortunate coincidence. The system's orbital period is very close to two years ($\sim$ 1.98 years), and the eclipse in the year 2016 occurred at a time when V2756~Sgr was not observable due to the solar conjunction. Later (2018, 2020, 2022), the beginnings of the eclipses were covered by the ASAS-SN observations at the very end of the observing season (just before the annual seasonal gap; Fig. \ref{fig:figure}B).

Interestingly, this is the very same reason why \citet{2013AcA....63..405G} did not report on the presence of eclipses in this system and why their value of the orbital period is different from ours. They used ASAS observations, and during the period covered by these data, most of the eclipses occurred during the seasonal gaps (Fig. \ref{fig:figure}A). Only one eclipse (2003) was partly covered (its second half; Fig. \ref{fig:figure}B), which was not sufficient to reveal the eclipsing nature of V2756~Sgr.

Besides the search for the eclipses in the available light curves, we also performed a period analysis of the data from DASCH ($g$ filter), ASAS-SN ($g$ filter), and ASAS ($V$ filter) using the ‘date compensated’ discrete Fourier transform \citep{1981AJ.....86..619F} in the Peranso software \citep{2016AN....337..239P}. All three datasets revealed a significant period of around 460 days, close to the value obtained by \citet{2013AcA....63..405G}. This variability might be connected with semi-regular pulsations of the cool component; however, it is rather long for typical giants in S-type symbiotic binaries. In addition, a period that is close to the orbital one was detected in all three light curves ($\sim$ 730 days in the DASCH data, slightly longer in ASAS-SN and ASAS datasets; Fig. \ref{fig:figure}C). 

\section{Gaia low-resolution spectroscopy}

Uncalibrated low-resolution BP/RP slitless spectra \citep[R $\sim$ 100; see more in][]{2016A&A...595A...1G,2022arXiv220800211G} are available directly at the GSA website for all transients alerted by \textit{Gaia}. Even though these data are neither calibrated to physical units nor wavelengths, they provide excellent information on the possible spectral variability, e.g., color changes during the detected brightenings/fadings. 

The observations for V2756~Sgr (see the GSA website for the spectra) typically show a prominent red continuum (TiO bands of the cool component are well detectable in the RP spectra even with the low-resolution of \textit{Gaia}), a hot continuum, and a strong H$\alpha$ emission line. Other emission lines are not detectable with such resolution. The data obtained during the last eclipse clearly show that the star became significantly redder. During the eclipse, the TiO bands became more prominent and were detectable even in the BP part of the spectrum. Such behavior confirms that the hot component of V2756~Sgr, together with the surrounding ionized nebula, is eclipsed by the cool component.

\section{Conclusions}
V2756~Sgr was previously not known as an eclipsing symbiotic system, but it was recently detected in an eclipse-like fading by \textit{Gaia}. Our investigation of the \textit{Gaia} photometric and spectroscopic data and the light curves from ASAS, ASAS-SN, and DASCH allowed us not only to confirm the eclipsing nature of this binary but also to refine its orbital period to 725 $\pm$ 3 days. By unlucky coincidence - the orbital period is very close to 2 years, and the eclipses occur around the time of the solar conjunction - not a single eclipse of this system has been observed from its beginning to the end until now. The approximate time of the next mid-eclipse of V2756~Sgr is JD\,$\sim$\,2\,460\,635 (November 20, 2024).



\begin{acknowledgments}
We acknowledge ESA Gaia, DPAC, and the Photometric Science Alerts Team. This research was supported by the \textit{Slovak Research and Development Agency} (APVV-20-0148).
\end{acknowledgments}

\bibliography{sample631}{}
\bibliographystyle{aasjournal}



\end{document}